# The Impact of Situation Clustering in Contextual-Bandit Algorithm for Context-Aware Recommender Systems

D. Bouneffouf
Télécom SudParis
9, rue Charles Fourier
91011 Evry, France
+33 1 60 76 47 82
Djallel.Bouneffouf@it-sudparis.eu

## Abstract
Most existing approaches in Context-Aware Recommender Systems (CRS) focus on recommending relevant items to users taking into account contextual information, such as time, location, or social aspects. However, few of them have considered the problem of user's content dynamicity. We introduce in this paper an algorithm that tackles the user's content dynamicity by modeling the CRS as a contextual bandit algorithm and by including a situation clustering algorithm to improve the precision of the CRS. Within a deliberately designed offline simulation framework, we conduct evaluations with real online event log data. The experimental results and detailed analysis reveal several important discoveries in context aware recommender system.

## Introduction
H.3.3 [**Information Search and Retrieval**]: *information filtering, Selection process, Relevance feedback.*

## General Terms
Algorithms

## 1. INTRODUCTION
A considerable amount of research has been done in recommending interesting content for mobile users. Earlier techniques in Context-Aware Recommender Systems (CRS) [3, 6, 13, 5, 23, 24] are based solely on the computational behavior of the user to model his interests regarding his surrounding environment like location, time and near people (the user's situation). The main limitation of such approaches is that they do not take into account the dynamicity of the user's content.

Few works found in the literature [14, 22] solve this problem by addressing it as a need for balancing exploration and exploitation studied in the "bandit algorithm" [21]. A bandit algorithm B exploits its past experience to select documents (arms) that appear more frequently. Besides, these seemingly optimal documents may in fact be suboptimal, because of the imprecision in B's knowledge. In order to avoid this undesired situation, B has to explore documents by choosing seemingly suboptimal documents so as to gather more information about them. Exploitation can decrease short-term user's satisfaction since some suboptimal documents may be chosen. However, obtaining information about the documents' average rewards (i.e., exploration) can refine B's estimate of the documents' rewards and in turn increases long-term user's satisfaction. Clearly, neither a purely exploring nor a purely exploiting algorithm works well, and a good tradeoff is needed.

The authors on [14, 22] describe a smart way to balance exploration and exploitation in the field of recommender systems. However, none of them consider the user's situation during the recommendation.

In order to give CRS the capability to provide the mobile user's information matching his/her situation and adapted to the evolution of his/her content (good exr/exp tradeoff in the bandit algorithm), we propose an algorithm witch takes into account the user's situation clusters for defining the (exr/exp) tradeoff, and then selects suitable situations for either exploration or exploitation.

The rest of the paper is organized as follows. Section 2 reviews some related works. Section 3 presents the user's model of our CRS. Section 4 describes the algorithms involved in the proposed approach. The experimental evaluation is illustrated in Section 5. The last section concludes the paper and points out possible directions for future work.

## 2. RELATED WORKS
We review in the following recent relevant recommendation techniques that tackle the two issues mentioned above, namely: following the evolution of the user's contents using bandit algorithm and considering the user's situation on recommender system.

### 2.1 Bandit Algorithms Overview
The (exr/exp) tradeoff was firstly studied in reinforcement learning in 1980's, and later flourished in other fields of machine learning [17, 20]. Very frequently used in reinforcement learning to study the (exr/exp) tradeoff, the multi-armed bandit problem was originally described by Robbins [21].

The ε-greedy is the one of the most used strategy to solve the bandit problem and was first described in [15]. The ε-greedy strategy choose a random document with epsilon-frequency (ε), and choose otherwise the document with the highest estimated mean, the estimation is based on the rewards observed thus far. ε must be in the open interval [0, 1] and its choice is left to the user.

Another variant of the ε-greedy strategy is what Cesa-Bianchi and Fisher [15] call the ε-decreasing strategy. In this strategy, the document with the highest estimated mean is always pulled except when a random document is pulled instead with an $\varepsilon_i$ frequency, where n is the index of the current round. The value of the decreasing $\varepsilon_i$ is given by $\varepsilon_i = \{\varepsilon_0/ i\}$ where $\varepsilon_0 \in ]0,1]$. Besides ε-decreasing, four other strategies are presented in [4]. Those strategies are not described here because the experiments done by [4] seem to show that, with carefully chosen parameters, ε-decreasing is always as good as the other strategies.

Compared to the standard multi-armed bandit problem with a fixed set of possible actions, in CRS, old documents may expire and new documents may frequently emerge. Therefore it may not be desirable to perform the exploration all at once at the beginning as in [9] or to decrease monotonically the effort on exploration as the decreasing strategy in [15].

Few research works are dedicated to study the contextual bandit problem on Recommender System, where they consider user's behavior as the context of the bandit problem.

In [10], authors extend the ε-greedy strategy by updating the exploration value ε dynamically. At each iteration, they run a sampling procedure to select a new ε from a finite set of candidates. The probabilities associated to the candidates are uniformly initialized and updated with the Exponentiated Gradient (EG) [11]. This updating rule increases the probability of a candidate ε if it leads to a user's click. Compared to both ε-beginning and decreasing strategy, this technique improves the results.

In [14], authors model the recommendation as a contextual bandit problem. They propose an approach in which a learning algorithm selects sequentially documents to serve users based on contextual information about the users and the documents. To maximize the total number of user's clicks, this work proposes the LINUCB algorithm that is computationally efficient.

The authors in [4, 9, 14, 15, 22] describe a smart way to balance exploration and exploitation. However, none of them consider the user's situation during the recommendation.

## 2.2 Managing the User's Situation

Few research works are dedicated to manage the user's situation on recommendation. In [7, 18] the authors propose a method which consists of building a dynamic user's profile based on time and user's experience. The user's preferences in the user's profile are weighted according to the situation (time, location) and the user's behavior. To model the evolution on the user's preferences according to his temporal situation in different periods, (like workday or vacations), the weighted association for the concepts in the user's profile is established for every new experience of the user. The user's activity combined with the user's profile are used together to filter and recommend relevant content.

Another work [13] describes a CRS operating on three dimensions of context that complement each other to get highly targeted. First, the CRS analyzes information such as clients' address books to estimate the level of social affinity among the users. Second, it combines social affinity with the spatiotemporal dimensions and the user's history in order to improve the quality of the recommendations.

In [3], the authors present a technique to perform user-based collaborative filtering. Each user's mobile device stores all explicit ratings made by its owner as well as ratings received from other users. Only users in spatiotemporal proximity are able to exchange ratings and they show how this provides a natural filtering based on social contexts.

Each work cited above tries to recommend interesting information to users on contextual situation; however they do not consider the evolution of the user's content.

| Recommender Systems | Evaluation Criteria | |
|---|---|---|
| | Following the evolution of user's content | Managing the user's situation |
| Bandit Algorithms [14,22] | ☺ | ☹ |
| Context-Aware Algorithms [3, 7, 13, 18] | ☹ | ☺ |
| **Clustering-ε-greedy** | ☻ | ☻ |

**Table 1: Recommendation algorithms comparison**

Table 1 compares the existing algorithms and the one we propose (Clustering-ε-greedy) w. r. t. two criteria, namely the evolution user's content and the consideration of the user's situation.

In our work we aim at taking advantage of the surveyed approaches and improve their weaknesses. As shown in Table 1, none of the mentioned works tackles both problems of the evolution user's content and user's situation consideration in the recommendation. This is precisely what we intend to do with our approach (Clustering-ε-greedy), exploiting the following new features:

- Modeling the CRS as a contextual bandit algorithm by considering the user's situation when managing the (exr/exp)-tradeoff on recommendation.
- Using clustering algorithm we accelerate the context process acquisition and we improve the precision of the CRS.

The two features cited above are not considered in the surveyed approaches as far as we know.

In what follows, we define briefly the structure of the user's model and the methods for inferring the recommendation situations. Then, we explain how to manage the exploration/exploitation strategy, according to the current situation.

## 3. USER AND CONTEXT MODELS

The user's model is structured as a case base, which is composed of a set of situations with their corresponding user's preferences, denoted $U = \{(S^i; UP^i)\}$, where $S^i$ is a user's situation (Section 3.2.1) and $UP^i$ its corresponding user's preferences (Section 3.1).

## 3.1 The User's Preferences

The user's preferences are contextual and might depend on many factors, like the location or the current task within an activity. Thus, they are associated to the user's situation and the user's activity. Preferences are deduced during the user's navigation activities. A navigation activity expresses the following sequence of events: (i) the user's logs in the system and navigates across documents to get the desired information; (ii) the user expresses his/her preferences about the visited documents. We assume that a visited document is relevant, and thus belongs to the user's preferences, if there are some observable user's behaviors through two types of preference:

- The direct preference: the user expresses his/her interest in the document by inserting a rate, like for example putting starts ("*") at the top of the document.

- The indirect preference: it is the information that we extract from the user's system interaction, for example the number of clicks on the visited documents or the time spent on a document.

Let *UP* be the preferences submitted by a specific user in the system at a given situation. Each document in *UP* is represented as a single vector $d=(c_1,...,c_n)$, where $c_i$ ($i=1, .., n$) is the value of a component characterizing the preferences of *d*. We consider the following components: the total number of clicks on *d*, the total time spent reading *d*, the number of times *d* was recommended, and the direct preference rate on *d*.

## 3.2 Context Model

A user's context *C* is a multi-ontology representation where each ontology corresponds to a context dimension $C=(O_{Location}, O_{Time}, O_{Social})$. Each dimension models and manages a context information type. We focus on these three dimensions since they cover all needed information. These ontologies are described in [1] and are not developed in this paper.

### 3.2.1 Situation Model

A situation is a projection on one or several user's context dimensions. In other words, we consider a situation as a triple $s = (O_{Location}.x_i, O_{Time}.x_j, O_{Social}.x_k)$ where $x_i$, $x_j$ and $x_k$ are ontology concepts or instances. Suppose the following data are sensed from the user's mobile phone: the GPS shows the latitude and longitude of a point "48.8925349, 2.2367939"; the local time is "Mon *May* 3 12:10:00 2012" and the calendar states "meeting with Paul Gerard". The corresponding situation is:

$S=(O_{Location}, "48.89,2.23",$
$O_{Time}."Mon\_May\_3\_12:10:00\_2012", O_{Social}. "Paul\_Gerard")$.

To build a more abstracted situation, we interpret the user's behavior from this low-level multimodal sensor data using ontologies reasoning means. For example, from *S*, we obtain the following situation:

*MeetingAtRestarant=($O_{Location}$.Restaurant, $O_{Time}$.Work_day, $O_{Social}$.Financial_client)*.

For simplification reasons, we adopt in the rest of the paper the following notation:

$S = (x_i, x_j, x_k)$. The previous example situation became thus:

*MeetingAtRestarant=(Restaurant, Work_day, Financial_client)*.

Among the set of captured situations, some of them are characterized as *high-level critical situations*.

## 4. THE PROPOSED RECOMMENDATION ALGORITHM

The problem of recommending documents can be naturally modeled as a multi-armed bandit problem with context information. In our case we consider the user's situation as the context information of the multi-armed bandit. Following previous work [12], we call it a contextual bandit. Formally, our contextual-bandit algorithm proceeds in discrete trials $t = 1...T$. For each trial *t*, the algorithm performs the following tasks:

**Task 1:** Let $S^t$ be the current user's situation, and *PS* the set of past situations. The system compares $S^t$ with the situations in *PS* in order to choose the most similar $S^p$ using the *RetrieveCase()* method (Section 4.2.1).

**Task 2:** Let *D* be the document collection and $D_p \in D$ the set of documents recommended in situation $S^p$. After retrieving $S^p$, the system observes the user's behavior when reading each document $d_i \in D_p$. Based on observed rewards, the algorithm chooses the document $d_p$ with the greater reward $r_p$ using the *RecommendDocuments*() method (Section 4.2.2).

**Task 3:** The algorithm improves its document-selection strategy with the new observation $(d_p, r_t)$. The updating of the case base is done using the *Auto_improvement()* method (Section 4.2.3), to accelerate the situation similarity computing this method includes a situation clustering algorithm.

Our goal is to design the bandit algorithm so that the expected total reward is maximized.

In the field of document recommendation, when a document is presented to the user and this one selects it by a click, a reward of 1 is incurred; otherwise, the reward is 0. With this definition of reward, the expected reward of a document is precisely its Click Through Rate (CTR). The CTR is the average number of clicks on a recommended document, computed dividing the total number of clicks on it by the number of times it was recommended. It is important to know here that no reward $r_{t,d}$ is observed for unchosen documents $d \neq d_t$ previously displayed.

## 4.1 The ε-greedy Algorithm

The ε-greedy strategy is sketched in Algorithm 1. For a given user's situation, the algorithm recommends a predefined number of documents, specified by parameter *N*. In this algorithm, $UC=\{d_1,...,d_P\}$ is the set of documents corresponding to the current user's preferences; $D=\{d_1,....,d_N\}$ is the set of documents to recommend; *getCTR* (Alg. 1, line 6) is the function which estimates the CTR of a given document; *Random* (Alg. 1, lines 5 and 8) is the function returning a random element from a given set; *q* is a random value uniformly distributed over [0, 1] which defines the exploration/exploitation tradeoff; *ε* is the probability of recommending a random exploratory document.

**Algorithm 1** The ε-greedy algorithm

1: **Input:** ε, UC, N
2: **Output:** D
3: D = ∅
4: **For** i =1 to i =N **do**
5:     q = Random({0,1})
6:     $d_i = \begin{cases} \arg\max_{UC-D}(getCTR(d)) & \text{if } q > \varepsilon \\ Random(UC) & \text{otherwise} \end{cases}$
7:
8:
9:     D = D ∪ $d_i$
10: **Endfor**

## 4.2 Clustering-ε-greedy()

To adapt the ε-greedy algorithm to a context aware environment, we propose to compute the similarity between the present situation and each one in the situation base; if there is a situation that can be reused; the algorithm retrieves it, and then applies the ε-greedy algorithm. The proposed Clustering-ε-greedy algorithm is described in Algorithm 6 and involves the following four methods:

### 4.2.1 RetrieveCase()

Given the current situation $S^t$, the *RetrieveCase()* method determines the expected user's preferences by comparing $S^t$ with the situations in past cases in order to choose the most similar one $S^p$. The method returns, then, the corresponding case ($S^p$, $UP^p$).

Let PS={ $S_{cl_1}^1, ..., S_{cl_n}^m$ } be the set of *m* past situations where {$cl_1, ..., cl_n$} is the set of clusters, and $S_{cl_i}^d$ is the centroid situation on the cluster $cl_{i\in\{0,...,n\}}$, and *n* the number of clusters. The clustering technique is described in Alg.5. The *RetrieveCase()* method detects from PS the cluster $cl_k$ including the most similar situation to the current one, by selecting the centroid situation $S_{cl_k}^d$ verifying Eq. 1(Alg.2), and then selecting, from the cluster $cl_k$, the situation verifying Eq. 2 (Alg.2).

**Algorithm 2** RetrieveCase()

1: **Input:** $S^t$, PS, UP
2: **Output:** $S^P$, $UP^p$
3: // select the nearest cluster
4: $S_{cl_k}^d = \arg\max_{S_{cl_i}^d, i \in (1,...,n)} \left( \sum_{j=1}^{3} \alpha_j \cdot sim_j(x_j^t, x_j^d) \right)$    (1)

// select the nearest situation on the cluster

5: $S_{cl_k}^p = \arg\max_{S_{cl_k}^c} \left( \sum_{j=1}^{3} \alpha_j \cdot sim_j(x_j^t, x_j^c) \right)$    (2)

6: Select $UP^p$ from UP

In Eq.1, $sim_j$ is the similarity metric related to dimension *j* between two concepts $x^t$ and $x^c$. This similarity depends on how closely $x^t$ and $x^c$ are related in the corresponding ontology (location, time or social). $\alpha_j$ is the weight associated to dimension *j*, and it is set out by using the arithmetic mean as follows:

$$\alpha_j = \frac{1}{T}\sum_{i=1}^{T} \gamma_j^i \quad (3)$$

In Eq. 3, $\gamma_j^i = sim_j(x_j^t, x_j^p)$ at trial $i \in \{1,...,T\}$ at the previous recommendation, where $x_j^p \in S^p$ and *T* the number of trials in the previous recommendation.

The similarity between two concepts of a dimension *j* in an ontological semantics depends on how closely they are related in the corresponding ontology (location, time or social). We use the same similarity measure as [25] defined by Eq. 4:

$$sim_j(x_j^t, x_j^c) = 2 * \frac{deph(LCS)}{(deph(x_j^t) + deph(x_j^c))} \quad (4)$$

In Eq. 2, LCS is the Least Common Subsumer of $x_j^t$ and $x_j^c$, and depth is the number of nodes in the path from the node to the ontology root.

### 4.2.2 RecommendDocuments()

In order to insure a better precision of the recommender results, the recommendation takes place only if the following condition is verified: $sim(S^t, S^p) \geq B$, where $sim(S^t, S^p) = \sum_j sim_j(x_j^t, x_j^p)$ and *B* the similarity threshold value.

To improve the adaptation of the ε-greedy algorithm to HLCS situations, we propose to make the following verification:

If the most similar $S^p \in HLCS$, the system recommends documents with a greedy strategy and inserts the current situation $S^t$ on the HLCS class of situations; otherwise the system uses the ε-greedy() method described in Alg. 1. Alg. 3 summarizes the functional steps of the *RecommendDocuments()* method.

**Algorithm 3** RecommendDocuments()

1. **Input:** ε, $UP^p$, $S^t$, $S^p$, N, B
2. **Output:** D
3. D = ∅
4. **If** $sim(S^t, S^p) \geq B$ **then**
5.     **If** $S^p \in HLCS$ **then**
6.        **For** i =1 to N **do**
7.           $d_i = \arg\max_{d \in (UP^p - D)} (getCTR(d))$
8.           D = D ∪ $d_i$
9.        **Endfor**
10.        HLCS ∪ $S^t$
11.     **Else** D = ε-greedy(ε, $UP^p$, N);
12.     **Endif**
13. **Endif**

### 4.2.3 Auto_improvement()

After recommending documents applying the *Recommend-Documents* method, two methods are used for improving the system:

- *UpdatePreferences (Alg.4)*: this method is used to update the user's preferences w. r. t. the number of clicks and number of recommendations for each recommended document on which the user clicked at least one time.
- *SituationClustering (Alg.5)*: this method is used to cluster the similar situations after a predefined number of recommendations.

#### 4.2.3.1 UpdatePreferences()

Depending on the similarity between the current situation $S^t$ and its most similar situation $S^p$ (computed with *RetrieveCase()*), being 3 the number of dimensions in the context, two scenarios are possible:

- $sim(S^t, S^p) \neq 3$: the current situation does not exist in the case base (Alg. 4, line 5); the *InsertCase()* method adds to the case base the new case composed of the current situation $S^t$ and the current user preferences $UP^t$.

- $sim(S^t, S^p) = 3$: the situation exists in the case base (Alg. 4, line 3); the *UpdateCase()* method updates the case having premise situation $S^p$ with the current user preferences $UP^t$.

| **Algorithm 4** UpdatePreferences() |
|---|
| 1: **Input:** $UP^t$, $S^t$, $S^p$     **Output:** Ø |
| 2: **if** $sim(S^t, S^p) = 3$ **then** |
| 3:     UpdateCase($S^p$, $UP^t$); |
| 4:     **else** InsertCase($S^t$, $UP^t$); |
| 5: **end if** |
| 6: **Endfor** |

#### 4.2.3.2 SituationClustering()

To accelerate the computing similarity between the current situation $S^t$ and the situations existing in the case base, we use one of the simplest and most efficient algorithms for clustering large datasets [10], the K-means algorithm, which is based on the analysis of variances. This algorithm clusters a group of situation vectors into a predefined number of clusters. It starts with random initial cluster centroids and keeps reassigning the situations in the dataset to cluster centroids based on the semantic similarity between the situations and the cluster centroid. The reassignment procedure will not stop until a fixed iteration number $t_{max}$ (Alg 5, line 6). We call this algorithm the *SituationClustering()* and it is shown in Alg. 5.

| **Algorithm 5** SituationClustering() |
|---|
| 1.  **Input:** $PS$, $N_c$, $t_{max}$, $ct$, |
| 2.  **Output:** $PS$ |
| 3.  **If** $tt = tt * ct$ **then** |
| //randomly initialize the $S^d_{cl_k}$ cluster centroid vectors |
| 1.    **For** $k=1$ to $N_c$ **do** $S^d_{cl_k}$ =Random($PS$); **Endfor** |
| //Assign each situation to the closest situation centroid of the cluster |
| 2.    **For** $t=1$ to $t_{max}$ **do** |
| 3.        **Foreach** $S^s_{cl_i}$ in $PS$ **do** |
| 4.        $S^d_{cl_t} = \arg\max_{S^d_{cl_k \in (1,\ldots,Nc)}} \left(sim(S^s_{cl_i}, S^d_{cl_k})\right)$ |
| 5.        **Assign** $S^s_{cl_i}$ **to the cluster** $cl_t$ |
| 6.    **Endforeach** |
| //Recalculate the cluster centroid vector $S^d$ |
| 7.    **Foreach** $k=1$ to $k=Nc$ **do** |
| 8.     $S^d_{cl_k} = \arg\max_{S^p_{cl_k}} (\frac{1}{n_{cl_k}} \sum_{S^e_{cl_k} \in (1,\ldots,n)} sim(S^p_{cl_k}, S^e_{cl_k}))$ |
| 9.    **Endforeach** |
| 10.   **Endfor** |
| 11.  **Endif** |

In Alg. 5, $Nc$ denotes the number of clusters, $n_{cl_k}$ the number of situations in cluster $cl_k$; $S^d_{cl_k}$ denotes the centroid vector of cluster $cl_k$; $S^p_{cl_k}$, $S^e_{cl_k}$ are sets of situations that belong to cluster $cl_k$, $S^s_{cl_i}$ is the set of situations that belong to cluster $cl_i$ and $tt$ is the parameter that activate the clustering algorithm if the iteration $tt$ of *Context-ε-greedy()* gets the value $ct$; this parameter is discussed in Section 5.2.1.

The Algorithm 6 summarizes the functional steps of the four methods described above:

**Algorithm 6** Clustering-ε-greedy()

1. **Input:** ε, N, PS, $S^t$, UP, B, $t_{max}$, ct, $N_c$,
2. **Output:** D
3. tt=0;
4. **For each** new situation S **do**
    // Retrieve the most similar case
5.    $(S^P, UP^p)$ = RetrieveCase($S^t$, PS,UP);
   // Recommend documents
6.    D=RecommendDocuments(ε,$UP^p$, $S^t$, $S^p$, N, B,);
7.    **Receive a feedback $UP^t$ from the user**
   // update user's profil
8.    Auto_improvement(PS, $UP^t$, $S^t$, $S^p$, $N_c$, $t_{max}$,ct, tt)
9.    tt=tt+1; //the number of iterations of the algorithm
10. **Endfor**

## 4.3 Finding the Optimal Exploration-Exploitation Trade-off

In order to set out the optimal trade-off value ε, we iteratively update it by the method *Compute-ε()* (Alg.7). First we assume that we have a finite number of candidate values for ε, denoted by Hε= ($ε_1$,..., $ε_T$).

Our goal is to select the optimal ε from Hε. To this end, we apply the ε-greedy algorithm for proposing an $ε_i$, and then we use a set of weights w = ($w_1$,...,$w_T$) to keep track of the feedback of each $ε_i$, $w_i$ is increased if we receive a number of clicks $l_i$ from the user when we use $ε_i$.

**Algorithm 7** Compute-ε ()

1. **Input:** Hε, N, PS, $S^t$, UP, B, $t_{max}$, ct, $N_c$, n
2. **Output:** Ø
3. εε= Ø, τ= 0, $w_i$ =1, i = 1,..., T
4. **For** t = 1 to n **do**
5.    τ =0.01* t
6.    q = Random({0,1})
7.    $ε_i$ = $\begin{cases} \text{select } i \text{ with } \arg\max_{(i)}(w_i) & \text{if } q \leq \tau \\ Random(Hε - εε) & \text{otherwise} \end{cases}$
8.    $A_i$= context-ε-greedy($ε_i$, N, PS, $S^t$, UP,B, ct, $t_{max}$, $N_c$);
9.    **Receive a click feedback $l_i$ from the user**
10.   $w_i$= $w_{i+}$ $l_i$;
11.   εε = εε ∪ $ε_i$
12. **Endfor**

In Alg. 7. εε is the set of ε that have been previously selected, n is the number of iteration of the learning algorithm, i is the identifier of ε and τ is the probability of proposing the argmax$_i$ ($w_i$), this parameter starts at low value and iteratively increases until the end of the learning step.

## 5. EXPERIMENTAL EVALUATION

In order to evaluate empirically the performance of our approach, and in the absence of a standard evaluation framework, we propose an evaluation framework based on a diary set of study entries. The main objectives of the experimental evaluation are: (1) to find the optimal parameters of our algorithm and (2) to evaluate the performance of the proposed algorithm w. r. t. the ε variation and the dataset size. In the following, we describe our experimental datasets and then present and discuss the obtained results.

### 5.1 Evaluation Framework

We have conducted a diary study with the collaboration of the French software company Nomalys[1]. This company provides a history application, which records the time, the current location, the social and navigation information of its users during their application use. The diary study has taken 18 months and has generated 178369 diary situation entries.

Table 2 illustrates three examples of such entries where each situation is identified by IDS.

| IDS | Users | Time | Place | Client |
|---|---|---|---|---|
| 1 | Paul | 11/05/2011 | 75060 Paris Cedex 02 | NATIXIS |
| 2 | Fabrice | 15/05/2011 | 2 rue Kellermann - 59100 Roubaix - France | MGET |
| 3 | Paul | 19/05/2011 | 90 Boulevard Pasteur, 75015 Paris | AMUNDI |

**Table 2: Diary situation entries**

Each diary situation entry represents the capture of contextual time, location and social information. For each entry, the captured data are replaced with more abstracted information using time, spatial and social ontologies. Table 3 illustrates three examples of such transformations.

| IDS | Users | Time | Place | Client |
|---|---|---|---|---|
| 1 | Paul | Workday | Paris | Finance client |
| 2 | Fabrice | Workday | Roubaix | Social client |
| 3 | John | Holiday | Paris | Telecom client |

**Table 3: Semantic diary situation**

From the diary study, we have obtained a total of 2759283 entries concerning the user's navigation, expressed with an average of 15.47 entries per situation. Table 4 illustrates examples of such diary navigation entries, where **Click** is the number of clicks on a document; **Time** is the time spent on reading a document, and **Interest** is the direct interest expressed by stars (the maximum number of stars is five).

| IdDoc | IDS | Click | Time | Interest |
|---|---|---|---|---|
| 1 | 1 | 2 | 2' | *** |
| 2 | 1 | 4 | 3' | * |
| 3 | 2 | 1 | 5' | * |

**Table 4: Diary navigation entries**

### 5.2 Finding the Optimal Parameters

In order to set out the optimal parameters of the *Clustering-ε-greedy* as the iteration of the clustering algorithm and the threshold similarity value, we use a manual classification as a baseline and compare it with the results obtained by our technique. So, we take a random sampling of 1783 situations which corresponds to 10% of the situations entries, we manually gather similar situations; then we compare the constructed groups with the results obtained by our algorithm, with different parameters of the situation clustering and the similarity methods.

---

[1] Nomalys is a company that provides a graphical application on Smartphones allowing users to access their company's data.

### 5.2.1 Parameterizing the SituationClustering algorithm

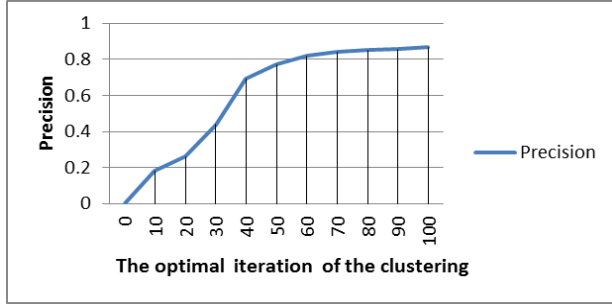

Figure 1.  Effect of the optimal itaration of clustering on the similarity precision

Figure 1 shows the effect of varying the iteration parameter $t_{max}$ (Alg. 5, line 3) in the interval [0, 100] on the overall precision. The results show that the precision has started to converge when $t$ has the value 60 achieving a precision of 0.819. Consequently, we use the identified optimal iteration value ($t_{max}$= 60) for testing our CRS.

In order to know how long our *Clustering-ε-greedy* can get optimal result without running the clustering method, we apply the *SituationClustering* at a different iteration of the *Clustering-ε-greedy* with an interval between [0, 100].

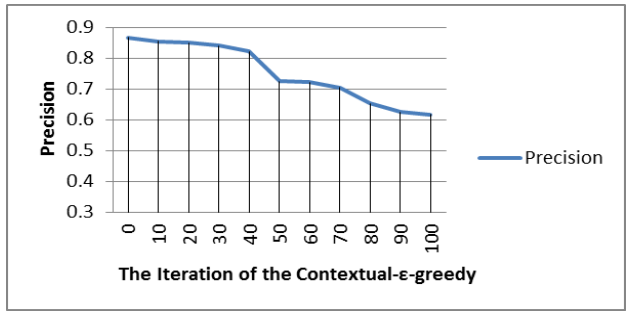

Figure 2.  Effect of the iteration of the *Clustering-ε-greedy* without using clustering on the similarity precision

Results in Figure 2 shows that the precision starts to decrease when *Clustering-ε-greedy* made a clustering after each 40 iterations $ct$=40 (alg. 5). Therefore, we run the clustering method (*SituationClustering*) with $ct$=40 and $t_{max}$= 60.

### 5.2.2 The threshold similarity value

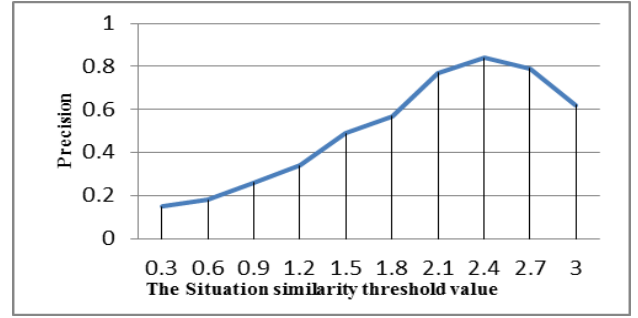

Figure 3.  Effect of B threshold value on the similarity precision

Figure 3 shows the effect of varying the threshold situation similarity parameter B (alg.3) in the interval [0, 3] on the overall precision. The results show that the best performance is obtained when B has the value 2.4 achieving a precision of 0.849. So, we use the identified optimal threshold value (B = 2.4) of the situation similarity measure for testing our CRS.

## 5.3 Experimental Results

In our experiments, we have firstly collected the 3000 situations (*HS*) with an occurrence greater than 100 to be statistically meaningful, and the 10000 documents (*HD*) that have been shown on any of these situations.

The testing step consists of evaluating the existing algorithms (*TestedAlgorithm* in Alg.8) by giving to these last as an entry a situation $S_j^t$ selected randomly from the sampling *HS*, where $t$ is the identifier of the situation, $j$ is the number of time that the situation $S^t$ has selected and $j_{max}$ is the number occurrence of $S^t$ in *HS*. The evaluation algorithm calculate and display the average CTR every 1000 iterations, The average CTR for a particular iteration is computed by the method *getAVCTR* (Alg. 8, line 7) and it is the ratio between the total number of clicks and the total number of displays. The number of documents (*N*) returned by the recommender system for each situation is 10 and we have run the simulation until the number of iterations *(i)* reaches 10000. The evaluation algorithm is described in Algorithm 8.

| **Algorithm 8** The Evaluation algorithm |
|---|
| 1. **Input:** *HS, HP, N, B, ct, Nc* |
| 2. **Output:** *AVCTR* |
| 3. $D = \emptyset, N=10, t_{max}=60, ct=40, B=2.4$ |
| 4. **For** $i =1$ **to** $10000$ **do** |
| 5. $S_j^t = Random(HS);$ |
| 6. $D_i = TestedAlgorithm(S_j^t, HD, B, N, t_{max}, ct, Nc);$ |
| 7. $D \cup D_i$ |
| 8. **If** $i = i*1000$ **then** $AVCTR= getAVCTR(D);$ |
| 9. $display(AVCTR);$ **Endif** |
| 10. **If** $j= j_{max}$ **then** *{HS}- $S_j^t$* **Endif** |
| 11. $j=j+1$ |
| 12. **Endfor** |

## 5.4 Results for ε Variation

In order to evaluate only the impact of considering the user's situation clustering in our bandit algorithm, we have compared the presented algorithm to a variant were we omitted the clustering part of the clustering-ε-greedy that we called ε-greedy. Each of the competing algorithms requires a single parameter ε. Figure 4 shows how the average CTR varies for each algorithm with the respective ε.

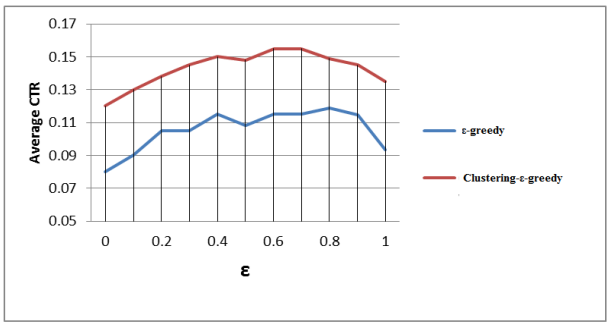

Figure 4. Variation ε tradeoff

Figure 4 shows that, when the parameter ε is too small, there is an insufficient exploration; consequently the algorithms have failed to identify interesting documents, and have got a smaller number of clicks (average CTR).

Moreover, when the parameter is too large, the algorithms seem to over-explore and thus lose a lot of opportunities to increase the number of clicks.

We can conclude from the evaluation that considering the user's situation clustering is indeed helpful for *Context-ε-greedy* to find a better match between the user's interest and the evolution of his content (documents).

## 6. CONCLUSION

In this paper, we have studied the problem of exploitation and exploration in context-aware recommender systems and propose a new approach that balances adaptively exr/exp by considering the situation clustering.

We have presented an evaluation protocol based on real mobile navigation contexts obtained from a diary study conducted with collaboration with the Nomalys French company. We have evaluated our approach according to the proposed evaluation protocol and show that it is effective.

In order to evaluate the performance of the proposed algorithm, we compare it with other standard exr/exp strategies. The experimental results demonstrate the positif impact of the situation clustering in the contextual bandit algorithm. In the future, we plan to extend our situation with more context dimension, and we plan to evaluate our approach using an online framework.